# Study on the composition optimization method for improving the fluidity of cast Ti$_2$AlNb alloy and its mechanism

Zhang Jian [a,b], Zhang Xiwen [c], Wang Hongwei [b,c], Wu Mengwei [a], Ma Xiong [b,c], Liang Xiaobo [b,c], Zhang Jianwei [b,c], Liu Ruiping [a], Sun Zhimin [c,d], Zhang Ji [b,c] a School of Electromechanical and Information Engineering, China University of Mining and Technology (Beijing), 100083, Beijing, China

b Gaona Aero Material Co., Ltd, 100081, Beijing, China

c Beijing Key Laboratory of New Superalloy Materials, 100081, Beijing, China

d Digital R&D Center of China Steel Research Group Co., Ltd., 100081, Beijing, China

*Abstract*: In this paper, the effects of Al, Nb main elements, Fe, Mo, W, Co, B, Si and their contents on the fluidity of Ti-22Al-25Nb alloy were investigated. The composition that was beneficial to improve the fluidity was screened through the thermodynamic software calculating thermophysical parameters affecting the fluidity of Ti$_2$AlNb alloy, the numerical simulation test of its fluidity and the verification test of the fluidity of optimized alloys. Finally, the improvement mechanism of the alloy fluidity was discussed. Results showed that the appropriate reduction of Nb element was better than Al element for the improvement of fluidity. The addition of trace Fe, B and Si elements were beneficial to the improvement of fluidity, the improvement effect of B element was best, while the addition of trace Mo, W, Co were not conducive to the improvement of fluidity. The cessation mechanism of Ti$_2$AlNb alloy is the cessation mechanism of the alloy with a wide crystallization temperature range. The composition which was most beneficial to improve the fluidity was Ti-22Al-24Nb-0.1B. The main reasons for the improvement of the fluidity had two sides: on the one hand, the reduction of 1at% Nb and the addition of 0.1at% B not only increased the superheat and crystallization latent heat of the alloy, but also reduced the melt viscosity and thermal conductivity, thus improving the fluidity. On the other hand, the TiB phase refined the grains, the fine grains prevented the dendrite from growing into developed dendrite networks, inhibited the adverse effect of the increase in the width of the solidification zone on the fluidity, reduced the flow resistance of the molten metal, and further improved the fluidity of the alloy.

# 1 Introduction

Nowadays, Ti$_2$AlNb alloy is regarded as one of the most promising lightweight high-temperature structural materials because of its low density, high specific strength and good oxidation resistance [1]. The traditional forging process produces Ti$_2$AlNb alloy products in face of low material utilization rate, complex machining process and large processing volume, which leads to high cost and long production cycle [2].. The preparation of Ti$_2$AlNb alloy castings by precision casting forming process can achieve near-net forming, reduce processing volume, shorten production cycle, and significantly reduce costs in the face of batch products. However, owing to the poor fluidity of the existing Ti-22Al-25Nb alloy, casting defects such as cold shuts, misrun and shrinkage are easy to occur when forming large thin-walled complex parts. Even if the external conditions such as shell preheating temperature, pouring temperature and centrifugal speed are adjusted to the limit of the equipment, it is still difficult to completely fill the mold. Therefore, it is necessary to find a solution to improve the fluidity of Ti$_2$AlNb alloy from the intrinsic factors that affect the fluidity.

Factors affecting the fluidity of the alloy include liquidus temperature, solidus temperature, solidification interval width, thermal conductivity, density, crystallization latent heat, dynamic viscosity, superheat, etc [3].. When external conditions are determined, these factors are mainly related to the alloy composition [4]. Therefore, adjusting the intrinsic factors affecting the fluidity through alloy composition design is one of the effective methods to improve the fluidity of Ti$_2$AlNb-based casting alloy. Previous research reports on Ti$_2$AlNb alloy alloying mainly focused on improving chemical or mechanical properties by alloying, such as adding Mo, Zr, W, Y, Ta, Si and other alloying elements to improve Ti$_2$AlNb alloy, such as oxidation resistance, compression property, tensile property or creep resistance [5-8], while the researchs on improving the fluidity of Ti$_2$AlNb alloy by alloying has not been reported. In cast titanium alloys, proper increase of Al content would improve the fluidity by increasing the latent heat of crystallization [9,10]. The addition of Fe would expand the solidification range and make the fluidity worse [11]. In a certain range, the addition of Si would greatly reduce the fluidity [12]. In cast TiAl alloy, adding appropriate amount of B, Si and Fe would increase the superheat, critical solid rate, solidification interval width or reduce the surface tension of the alloy, so as to improve the fluidity. The addition of Co and W would not significantly improve the fluidity of the alloy, while the addition of Mo would significantly increase the viscosity of the alloy and reduce the fluidity [13-17]. Among Ti$_2$AlNb alloys, Nb has the highest melting point. Proper reduction of Nb content might reduce the melting point of the alloy, increase the superheat, and improve the fluidity of the alloy. To sum up, due to the different solidification paths of different alloys, the effect and mechanism of elements on improving the fluidity of different alloys may be different.

This paper intends to take Ti-22Al-25Nb alloy as the base alloy (at.%, the same below), and systematically study the influence of Al, Nb main elements, Fe, Mo, W, Co, B and Si elements on the fluidity through Thermo-Calc thermodynamic calculation software and ProCAST casting simulation software, so as to screen the alloy composition beneficial to the improvement of the fluidity, and then carry out the verification test of the fluidity of the optimized alloy to further optimize the alloy composition, the effect mechanism of alloying elements and their contents on the fluidity are also discussed.

2 Materials and methods

In this paper, based on Ti-22Al-25Nb alloy, considering comprehensively the influence of main element content of Al and Nb on the mechanical properties of the alloy, the newly designed alloy components composition were Ti-(20~25) Al-(24~27) Nb, and (0~1) at.% trace alloying element was added, which include Fe, Mo, W, Co, B, Si. The specific alloy compositions were shown in Table 1.

The liquidus temperature, solidus temperature, thermal conductivity, density, enthalpy and dynamic viscosity of $Ti_2AlNb$ alloy at 0~2000 °C were calculated using TCTI3 and TCAL7 databases of Thermo-Calc software. Among them, thermal conductivity and density of the alloy were calculated using TCAL7 database, other thermophysical parameters were calculated using TCTI3 database.

Table 1 Summary of new alloy composition design

| Type | Composition(at.%) |
|---|---|
| Reference alloy | Ti-22Al-25Nb |
| Ti-xAl-25Nb alloy | Ti-(20~25)Al-25Nb |
| Ti-22Al-xNb alloy | Ti-22Al-(24~27)Nb |
| Microalloyed alloy | Ti-22Al-25Nb-(0.5, 1) Fe |
| | Ti-22Al-25Nb-(0.5, 1) Mo |
| | Ti-22Al-25Nb-(0.5, 1) W |
| | Ti-22Al-25Nb-(0.5, 1) Co |
| | Ti-22Al-25Nb-(0.5, 1) B |
| | Ti-22Al-25Nb-(0.5, 1) Si |

The 3D model of gating system and single spiral fluidity used in the numerical simulation test was shown in Fig. 1(a). Among them, the diameter of the cross section of the channel was 8 mm. The model was imported into ProCAST software for grid division, and then shell and virtual sandbox were added in the process, the results were shown in Fig. 1(b), (c) and (d). The thickness of the shell was 10mm and the size of the virtual sand box was 500×500×300 mm. There were 85538 surface grids and 849478 volume grids. The alloy components in Table 1 were used for the pouring system and casting materials, and the thermal physical parameters used for simulation were calculated by Thermo-Calc software. The shell was made of mullite refractory and the sand box was made of silica sand. The parameters related to the shell and the sand box were obtained from ProCAST commercial databases. The casting temperature was 2000 °C, the filling time was 3 s, and the preheating temperature of the shell was 200 °C. The heat transfer coefficient between the casting and the shell was 500 W/(m$^2$·K), and the heat transfer coefficient between shell and sand box was 300 W/(m$^2$·K). The cooling mode of sand box, castings and exposed parts of mold shell was air cooling. The filling form was gravity filling, the reference pressure was 5 Pa, and the filling parameter was 50. After the simulation operation parameters were set, flow simulation tests of Ti$_2$AlNb alloy with different components were carried out immediately. After the test, the filling length of each alloy was recorded and compared, and then alloy components used in the flow verification test were selected according to their flow improvement effect.

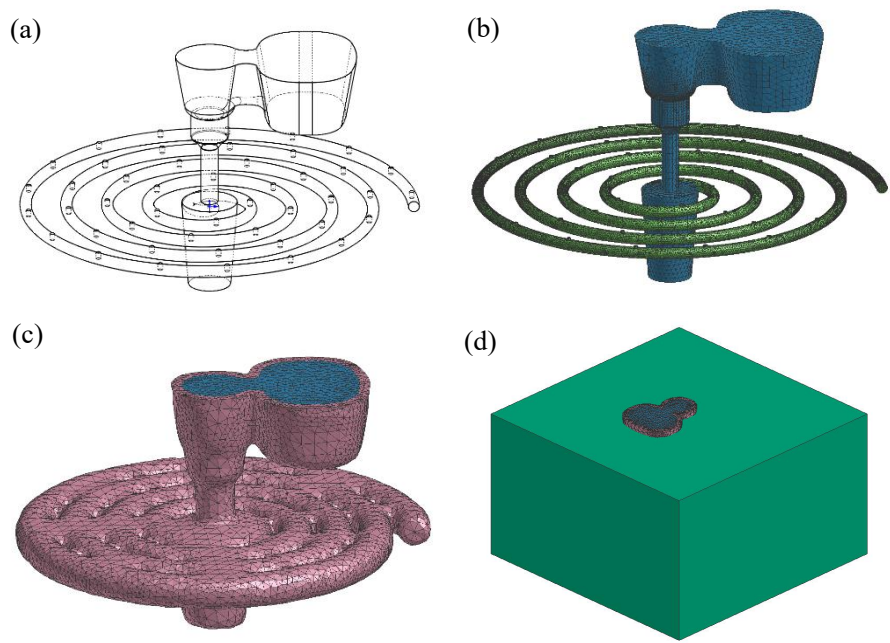

Fig. 1 The 3D model used in the fluidity simulation test and its meshing results
(a) Single spiral flow sample; (b), (c), (d) grid division results from ProCAST software

The parent alloy ingot of optimum alloy was prepared by vacuum induction suspension furnace melting, and then triple helical flow sample (see Fig. 2(a)) was poured by secondary melting in the suspension furnace. In order to ensure the consistency and stability of the pouring effect, the trihelix graphite casting mould was selected for the pouring test, as shown in Fig. 2(b). The gravity static mode was used for pouring. After pouring, the filling lengths of three flow channels were recorded, and their average values were taken to compare the improvement effect of the fluidity. The wire cutting method was used to cut the cross sections of different test alloy channels at different positions respectively to obtain metallographic samples (see Fig. 3). The samples were subjected to chemical corrosion after fine grinding by sandpaper and electrolytic polishing. The macroscopic microstructure of the samples was observed by DSX1000 stereocscope, and then metallographic microscope was used to observe the cross section microstructure of each channel at 5mm away from sprue. The corrosive agent ratio was HF: $HNO_3$: $H_2O$ = 2:3:7. The phase type of borides were determined by selective electron diffraction pattern (SAED) using F20 transmission electron microscopy (TEM).

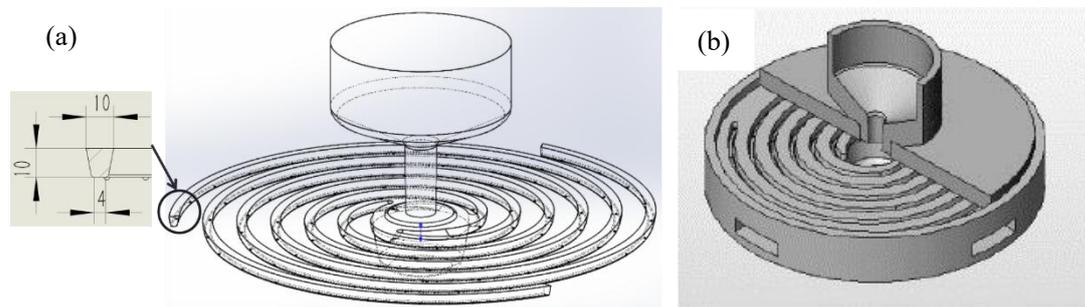

Fig. 2 3D model and graphite casting used in fluidity verification test

(a) trihelix flow sample; (b) graphite casting mould

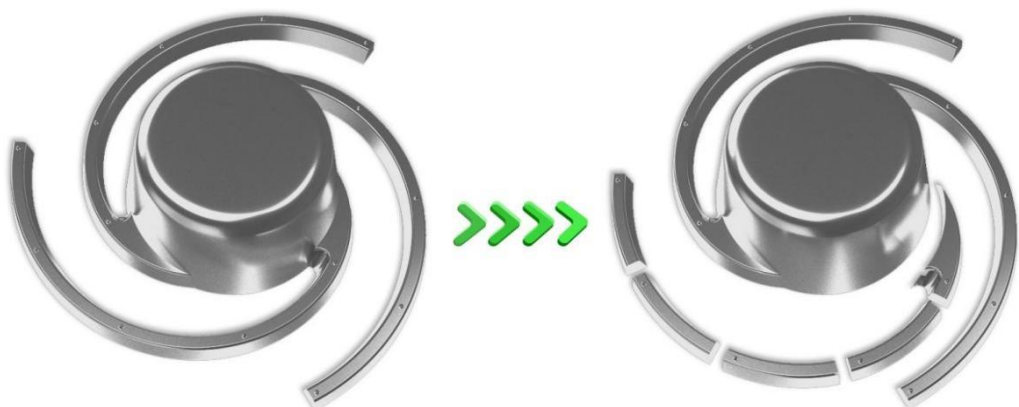

Fig. 3 Schematic diagram of metallographic sample sampling

## 3 Results

*3.1 Influence rules of alloying elements on thermal physical parameters*

Factors affecting the fluidity of the alloy include the liquidus temperature, solidus temperature, solidification interval width, thermal conductivity, density, crystallization latent heat, dynamic viscosity, superheat, etc [3]. According to the effects of various thermophysical parameters on the fluidity, the description is as follows: (1) Superheat: when the pouring temperature is constant, the decrease of the liquidus temperature of the alloy is equivalent to the increase of the superheat of the alloy, it will reduce the melt viscosity, and improve the fluidity[13]. (2) Solidification interval width: the width of solidification interval decreases, the time of melt in liquid becomes longer, and the fluidity increases[16]. (3) Thermal conductivity: the smaller the thermal conductivity, the smaller the liquid-solid two-phase zone during solidification, the smaller the flow resistance, and the better the fluidity. (4) Viscosity: in the early stage of mold filling, the effect of melt viscosity on the fluidity is small under turbulent conditions. At the later stage of mold filling, the melt viscosity has a greater impact on the fluidity under laminar flow state, the viscosity increases, the melt flow resistance increases, and the alloy fluidity decreases[18]. (5) Density: when the specific heat capacity is fixed, the higher the density of the alloy, because it contains a large amount of heat, the longer it stays in the liquid state under the same superheat, the better the fluidity of the alloy. (6) Crystallization latent heat: the more crystallization latent heat releases during solidification, the slower the solidification process, and the better the fluidity [4, 16]. In this paper, the enthalpy method was used to treat the latent heat of crystallization, and the enthalpy H (T) was introduced:

$$H(T) = \int_{T_f}^{T} c_p(T)\, dT = c_p(T - T_f) + L, T \geq T_f \qquad \text{Formula (1)}$$

Where, $c_p$ is the specific heat capacity, $L$ is the latent heat of crystallization, $T_f$ is the phase transition temperature [19]. Since the latent heat of crystallization mainly plays a role in solidification, the effect of the latent heat of crystallization in the solid-liquid two-phase region on the fluidity of the alloy is only discussed. $\langle c_p \rangle \approx \frac{dH}{dT}$, that is, the first derivative of enthalpy versus temperature is expressed as $\langle c_p \rangle$. Let $T_f$ is the temperature at which the liquid phase of an alloy completely transforms into a solid phase, and $T$ is the temperature at which the liquid phase begins to transform

into a solid phase. Assumption $c_p$ is constant, there are:

$$\langle c_p \rangle \approx \frac{H(T)-H(T_f)}{T-T_f} \qquad \text{Formula (2)}$$

Substituting Formula (2) into Formula (1), then:

$$L = H(T_f) \qquad \text{Formula (3)}$$

That is, the latent heat of crystallization is equal to the corresponding enthalpy when the liquid phase is completely transformed into a solid phase. When analyzing the influence of crystallization latent heat on the fluidity of the alloy, the curve of enthalpy changing with temperature in the liquid-solid two-phase region was approximately regarded as a straight line. In addition, density was not the main internal factor affecting alloy fluidity [4], so the effect of density on fluidity was not considered.

The influence of alloy composition on the fluidity is mainly caused by different composition and different crystallization characteristics of alloy [4]. The crystallization characteristic of the alloy depends on the solidification interval width [20], and the influence of the alloy composition on the fluidity mainly depends on the solidification interval width. Since the cessation mechanism of Ti$_2$AlNb alloy melt has no published literature for reference, it was impossible to judge the effect of solidification interval width on the fluidity of the alloy. Referring to the cessation mechanism mechanism of titanium alloy and TiAl alloy [16, 20], it was assumed that Ti$_2$AlNb alloy has the stopping flow mechanism characteristics of alloy with narrow crystallization temperature range. In addition, the discussion of the influence of solidification interval width on the flow is generally carried out on the premise that the superheat was constant [4]. It was inferred that the influence of superheat should be considered when discussing the influence of solidification interval width on the melt. Therefore, this part mainly focused on the influence on the solidification zone width and superheat of the alloy, and the influence of other factors would not be discussed temporarily.

Thermo-Calc thermodynamic calculation software was used to calculate the thermophysical parameters of alloy components in Table 1, and the results of liquidus/solidus temperature and solidification interval width of each alloy could be obtained directly. The influence of the main elements Al and Nb on the liquidus/solidus temperature of Ti$_2$AlNb alloy was shown in Fig. 4. From Fig. 4 (a) and (b), it could be seen that increasing the Al content and decreasing the Nb content basically did not change the width of solidification zone. It could be seen from Fig. 4(a) that the liquidus temperature of Ti-xAl-25Nb alloy decreased with the increase of Al content, increasing by 1at.% Al, and the liquidus temperature of the alloy decreased by about 2.26 °C. It could be seen from Fig. 4(b) that the liquidus temperature of Ti-22Al-xNb alloy decreased with the decrease of Nb content, which decreased by 1 at.% Nb, and the liquidus temperature of the alloy decreased by about 7.63 °C. It was obvious that compared with increasing 1at% Al, decreasing 1at% Nb, the liquidus temperature decreased more, so the degree of superheat increased more, which was more beneficial to fluidity. Although reducing Nb content was beneficial to improve the fluidity of $Ti_2AlNb$ alloy to a certain extent, excessive reduction of Nb content would deteriorate the mechanical properties of the alloy such as tensile properties, creep resistance and fracture toughness [19]. Considering the influence of casting properties (i.e. fluidity) and mechanical properties of the alloy, it was proposed to reduce 1 at% Nb on the basis of Ti-22Al-25Nb alloy, and set the alloy composition adjusted as Ti-22Al-24Nb alloy by main elements.

The effect of trace alloying elements such as Fe, Mo, W, Co on the liquidus/solidus temperature of Ti-22Al-25Nb alloy was shown in Fig. 5. It could be seen from Fig. 5 that the liquidus temperature of Ti-22Al-25Nb alloy increased with the increase of Mo and W content, with the maximum increment of 35.4 and 33.4 °C respectively; The liquidus temperature of Ti-22Al-25Nb alloy decreased with the increase of Fe and Co content, and the maximum reduction was 14.9 °C and 17.8 °C respectively. The solidification zone width of Ti-22Al-25Nb alloy increased with the increase of Mo, W, Fe and Co content, with the maximum increment of 24.3 °C, 25.0 °C, 41.1 °C and 304.0 °C, respectively. W and Mo elements increased the liquidus temperature of the alloy, and increased the width of solidification zone, which was unfavorable to the fluidity, so these two elements would not be considered in the subsequent calculation. Increasing the content of Fe could increase the solidification zone width and superheat of the alloy, which may be beneficial to fluidity. Although increasing the content of Co could improve the superheat of the alloy, the width of its solidification zone would also increase sharply, which was likely to offset the beneficial effect of increasing the superheat on the fluidity of the alloy and worsen its fluidity, so the effect of Co on the fluidity of the alloy was not considered.

The effect of trace alloying elements B and Si on the liquidus/solidus temperature of Ti-22Al-25Nb alloy was shown in Fig. 6. It could be seen from Fig. 6 that the liquidus temperature of Ti-22Al-25Nb alloy decreased with the increase of B and Si content, with the maximum reduction of 14.6 and 15.7 °C respectively. The solidification zone width of Ti-22Al-25Nb alloy increased at first and then remained unchanged with the increase of B content, with the maximum increment of 89.2 °C; The solidification zone width of Ti-22Al-25Nb alloy increased with the increase of Si content, and the maximum increment was 89.3 °C. It could be seen that the liquidus temperature of the alloy decreased with the addition of B and Si, but the width of the solidification zone increased. The influence of these two elements on the fluidity needed further study.

In view of the fact that adding a small amount of B and Si elements in titanium alloy and TiAl alloy can refine the grains [21, 22], the fine grains will hinder the growth of dendrites, avoid the formation of solid dendrite network, reduce the melt flow resistance, and improve the fluidity of the alloy [23]. The research results of literature [24] shown that adding trace B element into Ti-22Al-20Nb-2W alloy could refine its B2 phase grain. Based on the above analysis, after adding a small amount of B and Si elements into Ti-22Al-25Nb alloy, the beneficial effects of increasing the superheat of the alloy and refining the grain on the fluidity may offset the adverse effects of increasing the width of the solidification zone on the fluidity, so the influence of B and Si elements on the fluidity of the alloy was considered. When adding alloy elements to improve the fluidity, the effect of the addition amount on the mechanical properties of the alloy should also be considered comprehensively. The addition of (0.004~0.06) at% Fe to Ti-22Al-24Nb-0.5Mo alloy would not significantly affect the tensile properties of the alloy [25]. After adding 0.2at% B to Ti-22Al-20Nb-2W alloy, a large amount of TiB phase precipitated at the grain boundary, which refined the B2 phase grain and improved the ductility of the alloy at room temperature [26]. After adding 0.9 at% Si to Ti-24Al-14Nb-3V-0.5Mo alloy, 0.03% vol $Ti_5Si_3$ particles would precipitate in the B2 phase matrix, which destroyed the continuity of the B2 matrix, and caused large strain mismatch between the silicide and the matrix interface, then reduced the room temperature ductility of the alloy [24]. To sum up, on the basis of Ti-22Al-25Nb alloy, reducing the content of 1at% Nb and adding trace elements of Fe, B and Si were expected to improve the fluidity of the alloy. However, considering the mechanical properties of the casting alloy, the addition amount of trace alloying elements was limited to (0~0.5) at% in this paper. Subsequently, ProCast numerical simulation would be used to simulate the casting filling effect based on the following alloy components: Ti-22Al-25Nb, Ti-22Al-24Nb,

Ti-22Al-25Nb - (0~0.5) X (X was Fe, B and Si).

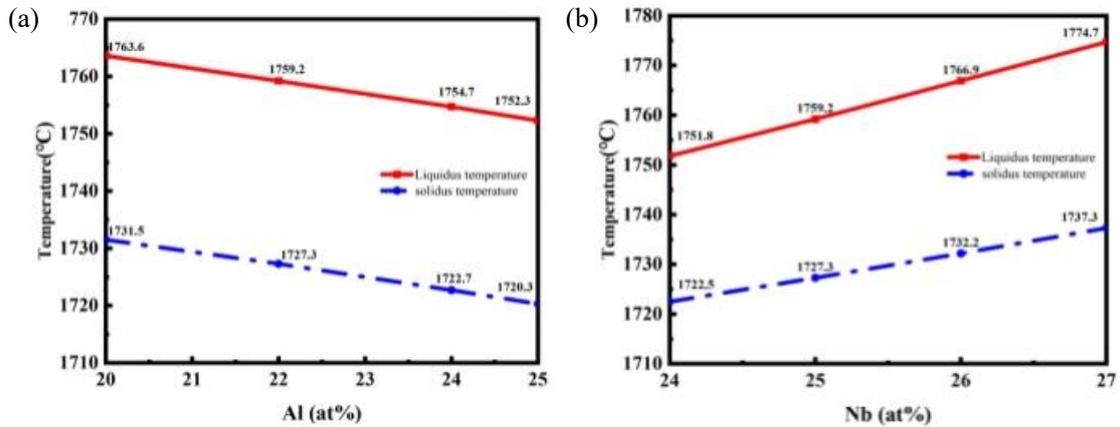

Fig. 4 Effect of main elements Al and Nb on liquidus/solidus temperature of Ti$_2$AlNb alloys

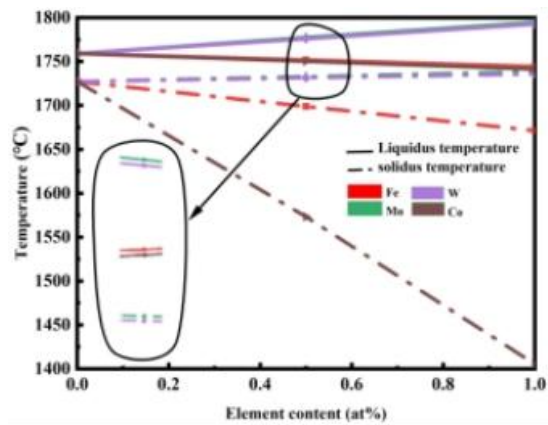

Fig. 5 Effect of trace alloying elements Fe, Mo, W and Co on the liquidus/solidus temperature of Ti-22Al-25Nb alloy

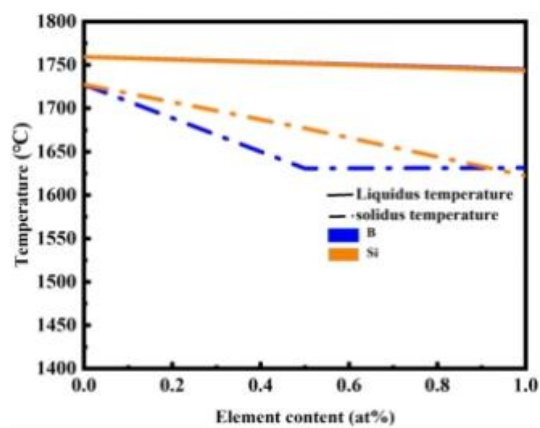

Fig. 6 Effect of trace alloying elements B and Si on the liquidus/solidus temperature of Ti-22Al-25Nb alloy

*3.2 Numerical simulation study of fluidity improvement effect*

Taking the alloy components selected in Section 3.1 as the test materials, the liquidus/solidus temperature, thermal conductivity, density and enthalpy of each alloy were calculated by Thermo-Calc software in the temperature range of 0~2000 °C, then input them into the ProCast numerical simulation software. The casting filling simulation was carried out using the single spiral flow mold in Fig.1. The filling results of Ti-22Al-25Nb alloy and Ti-22Al-24Nb alloy were shown in Fig.7, the filling results of microalloyed $Ti_2AlNb$ alloys were shown in Fig.8.

It could be seen from Fig.7 that the filling length of Ti-22Al-25Nb alloy was 1220 mm, and the filling length of Ti-22Al-24Nb alloy obtained by reducing 1 at% Nb on the basis of that was 1240mm, which was 20 mm more than that of the base alloy(+1.6%). It could be seen that the reduction of 1at% Nb prolonged the filling length and improved the fluidity of melt, which was consistent with the analysis result of adjusting the Al/Nb main element to improve the fluidity in Section 3.1.

The change of filling length with the content of Fe, Si and B elements was shown in Fig.8. It could be seen from Fig.8 that on the basis of Ti-22Al-25Nb alloy, the filling length had an increasing trend with the increase of Fe content. When the Fe content was 0.1 at%, the improvement effect of filling length was the most obvious, the filling length increased by 25mm (+2.0%). When the Fe content was 0.2 at%, the improvement effect of filling length decreased slightly, the filling length increases by 20mm (+1.6%) on the basis of the original alloy. As the Fe content continued to increase to 0.5at%, the alloy filling length did not increase, which was basically consistent with the 0.2at% Fe alloy. Increasing the content of Si element could also improve the filling effect. When the content of Si was 0.1~0.3at%, the filling length of the alloy increased by 20 mm (+1.6%), the filling improvement effect decreased. When the content of Si was 0.5at%, the filling length only increased by 10 mm (+0.8%). At the same time, excessive addition of Si content would increase the proportion of forming $Ti_5Si_3$ brittle phase, which was unfavorable to plasticity. With the increase of B content, the filling length of the alloy first increased and then decreased. When B content was 0.1at%, the filling length increased by 55 mm (+4.5%). With the increase of B content from 0.2at% to 0.4at%, the increase of filling length decreased from 45mm to 30mm. When the content of B increased to 0.5at%, the filling length decreased by 30mm. In general, the improvement effect of adding B element on the basis of Ti-22Al-25Nb alloy was better than that of adding Fe and Si elements and the effect of reducing 1at% Nb element. To sum up, the addition of 0.1at% B had the best effect on the filling improvement. Combining the reduction of 1at% Nb and the addition of trace B elements may further improve the fluidity of melt.

Combined with the effect of reducing Nb content and adding trace B element on the fluidity, new alloy composition Ti-22Al-24Nb (0.1~0.5) B were designed. The same thermodynamic calculation and numerical simulation method were used to study the effect of improving the fluidity of the new alloy. The results of filling length comparison were shown in Fig.8. On the basis of Ti-22Al-24Nb alloy, the addition of B element further improved the filling effect. The change rule of filling length with B content was similar to that of Ti-22Al-25Nb alloy. With the increase of B content, the filling length of the alloy first increased and then decreased. The addition of 0.1at% B and 0.2at% B alloys had the largest increase in filling length, which was 75mm (+6.1%) higher than the original alloy.

The numerical simulation study on the pouring and filling effect of $Ti_2AlNb$ alloy through ProCast software shown that reducing 1% Nb and adding 0.1~0.2at% B elements had the best effect on improving the fluidity, which was better than the effect of Fe and Si elements. The optimized alloy composition was Ti-22Al-24Nb-0.1B and Ti-22Al-24Nb-0.2B. In the next step, the alloy pouring test would be carried out to verify the actual improvement effect of Ti-22Al-24Nb-0.1B alloy and Ti-22Al-24Nb-0.2B alloy on the fluidity.

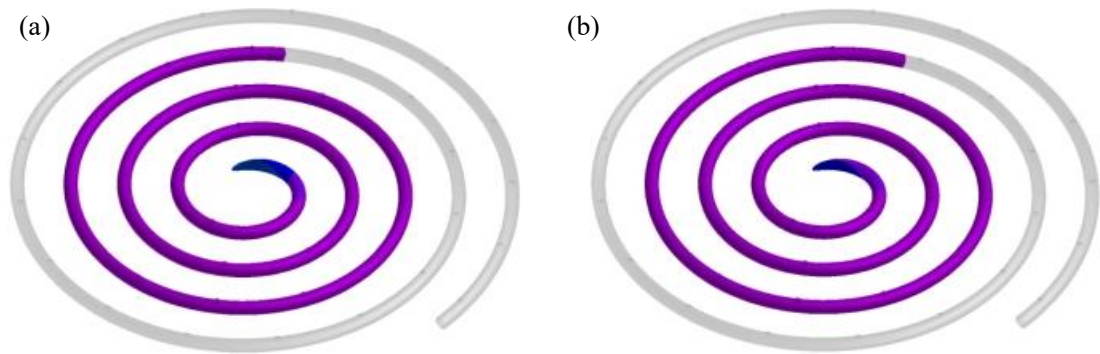

Fig. 7 Numerical simulation results of fluidity of Ti-22Al-25Nb alloy and Ti-22Al-24Nb alloy

(a) Ti-22Al-25Nb alloy; (b) Ti-22Al-24Nb alloy

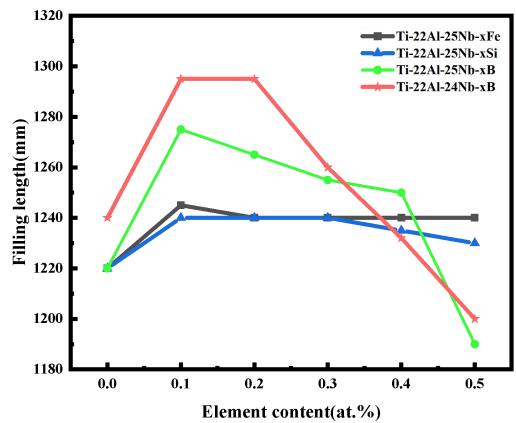

Fig. 8 Numerical simulation results of fluidity of microalloyed Ti$_2$AlNb alloys

*3.3 Research on fluidity verification test of optimized alloy*

In order to verify the actual improvement effect of the fluidity of the selected alloys, the ingots of three Ti$_2$AlNb alloy selected in Section 3.2 were prepared by melting and used to carry out the fluidity verification test study. The measured results of the three alloy compositions were basically consistent with the nominal compositions (seen Table 2). The relevant results of the liquidity validation test study were shown in Fig.9 and Table 3. Fig.9 shown the three spiral flow samples of the three alloys. According to the flow samples, the actual filling length of the three alloys was calculated (see Table 3). It should be pointed out here that the filling section of the three-screw flow sample model was smaller than that of the single-screw flow sample model. At the same time, the actual pouring conditions would be different from the ideal conditions of the simulation calculation, so the filling length would be different from the simulation calculation results. It could be seen from Table 3 that the average filling length of Ti-22Al-25Nb alloy was 228mm, and the average filling length of Ti-22Al-24Nb-0.1B alloy was 240mm, which was 5.3% more than that of the base alloy. The test verification results were consistent with the simulation results, which shown that on the basis of the original alloy, the reduction of 1at% Nb and the addition of 0.1at% B had obvious effect on improving the fluidity of the alloy. However, the average filling length of Ti-22Al-24Nb-0.2B alloy was only 154mm, it was 32.5% shorter than that of the base alloy, and the fluidity was deteriorated, which was contrary to the improvement rule of fluidity in the simulation results.

Table 2 Comparison of nominal composition and measured composition of three Ti$_2$AlNb alloys

| Nominal composition (at%) | Measured composition (at%) |
| --- | --- |
| Ti-22Al-25Nb | Ti-21.98Al-25.39Nb |
| Ti-22Al-24Nb-0.1B | Ti-21.21Al-24.41Nb-0.10B |
| Ti-22Al-24Nb-0.2B | Ti-22.50Al-24.42Nb-0.19B |

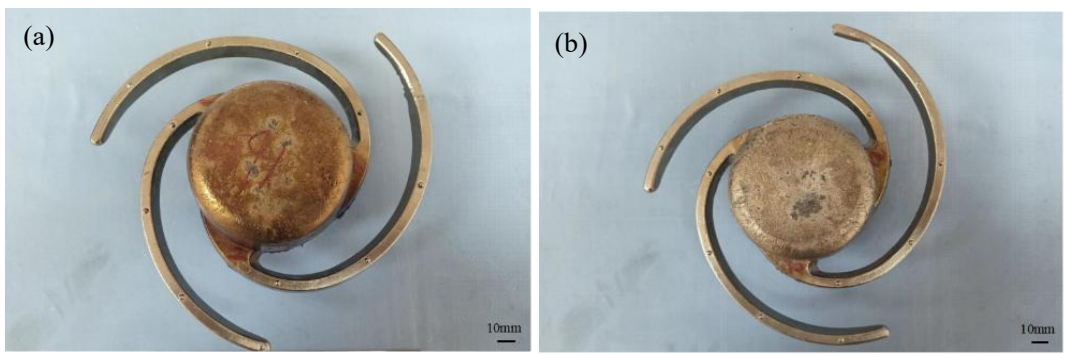

(a)　　　　　　　　　(b)

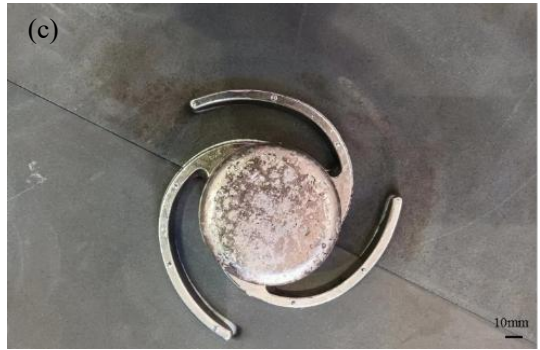

Fig.9 Fluidity verification test results of thereTi$_2$AlNb alloys

(a) Ti-22Al-25Nb alloy; (b) Ti-22Al-24Nb-0.1B alloy; (c) Ti-22Al-24Nb-0.2B alloy

Table 3 Actual filling length of three Ti$_2$AlNb alloys (mm)

| Alloy composition (at%) | Channel 1 | Channel 2 | Channel 3 | Average value | Increment (%) |
|---|---|---|---|---|---|
| Ti-22Al-25Nb | 205 | 241 | 238 | 228 | 0 |
| Ti-22Al-24Nb-0.1B | 228 | 263 | 229 | 240 | 5.3 |
| Ti-22Al-24Nb-0.2B | 162 | 150 | 150 | 154 | -32.5 |

## 4 Discussion

*4.1 Cessation mechanism of Ti$_2$AlNb alloy melt flow*

In order to explore the mechanism of improving the fluidity of Ti-22Al-24Nb-0.1B alloy, the cessation mechanism of Ti$_2$AlNb alloy melt flow was first discussed, and then the mechanism of improving the fluidity was analyzed by comparing several main factors affecting the fluidity, such as the width of solidification interval, superheat and crystallization latent heat.

At present, no studies have reported the stopping flow mechanism of Ti$_2$AlNb alloy. M C. Flemings taken Al-Sn alloy as the research object, proposed the stopping flow mechanism of alloy melt in narrow crystallization temperature range and wide crystallization temperature range [27]. The macro structure of the alloy in the narrow crystallization temperature range is columnar crystal. The mechanism of melt stopping flow is that the columnar crystal growing from the mold wall to the center at a certain position before the end contacts with each other, and the melt flow channel is blocked; the macro structure of the alloy with a wide crystallization temperature range is equiaxed crystal, the farther away from the sprue, the finer the grain. The mechanism of the melt stopping flow is that the front end of the liquid flow cools the fastest and crystallizes first. After the number of crystals reaches the critical solid rate, a continuous dendrite network is formed, and the melt flow channel is blocked [4].

Some literatures had reported the stopping flow mechanism of titanium alloy and Ti-Al intermetallic compound. Yang Chuang et al. [20] founded that the main reason why the Ti-6Al-4V alloy melt stopped flowing was that the columnar crystal at the root was rich in growth, and the formation of transgranular structure led to necking, it prevented the flow of liquid, making the alloy melt stop flowing. The Ti-6Al-4V alloy conformed to the characteristics of the stop flow mechanism in a narrow crystallization temperature range. Jianchao Han et al. [16] founded that the stopping flow mechanism of the alloy melt was similar to that of Ti-6Al-4V alloy in the process of studying the effect of $TiB_2$ on the microstructure and fluidity of Ti-48Al-2Cr-2Nb alloy, and the mechanism of end necking plugging was the main one, and the mechanism of front end plugging was the auxiliary one, which was characterized by the stopping flow mechanism of alloy with narrow crystallization temperature range.

In this study, the macrostructure of the cross section of the three components $Ti_2AlNb$ alloy at different positions from the sprue was shown in Table 4. It could be seen from Table 4 that the macrostructure of the cross-section of Ti-22Al-25Nb alloy at different positions from the sprue was equiaxed crystal structure, which was significantly different from Ti-6Al-4V alloy and Ti-48Al-2Cr-2Nb alloy. The grain size at different positions in the mold filling runner was measured by the method of cutting line. The results shown that the grain size tends to decrease with the distance from the sprue. The average grain size at the nearest position (about 50 mm) to the sprue was 242 μm. The average grain size at the farthest position (about 200 mm) from the sprue was 217 μm. It could be seen that the macroscopic structure characteristics of Ti-22Al-25Nb alloy were consistent with the characteristics of alloy stop flow mechanism in a wide crystallization temperature range. Comparing and observing the macrostructure of Ti-22Al-24Nb-0.1B and Ti-22Al-24Nb-0.2B alloys at different positions from the sprue, it could be seen that, like the base alloy, the macrostructure of the two B-containing alloys were also equiaxed, and the grains also shown the characteristics of the smaller average grain size as the farther away from the sprue (see Figure 10). On the basis of the original alloy, reducing 1at.% Nb and adding trace amount of B did not change the characteristics of the stopping flow mechanism of the alloy.

Table 4 Macro structure of cross section at different positions from the sprue

| composition (at%) \ Distance from the sprue (mm) | 50 | 100 | 150 | 200 | 250 |
|---|---|---|---|---|---|

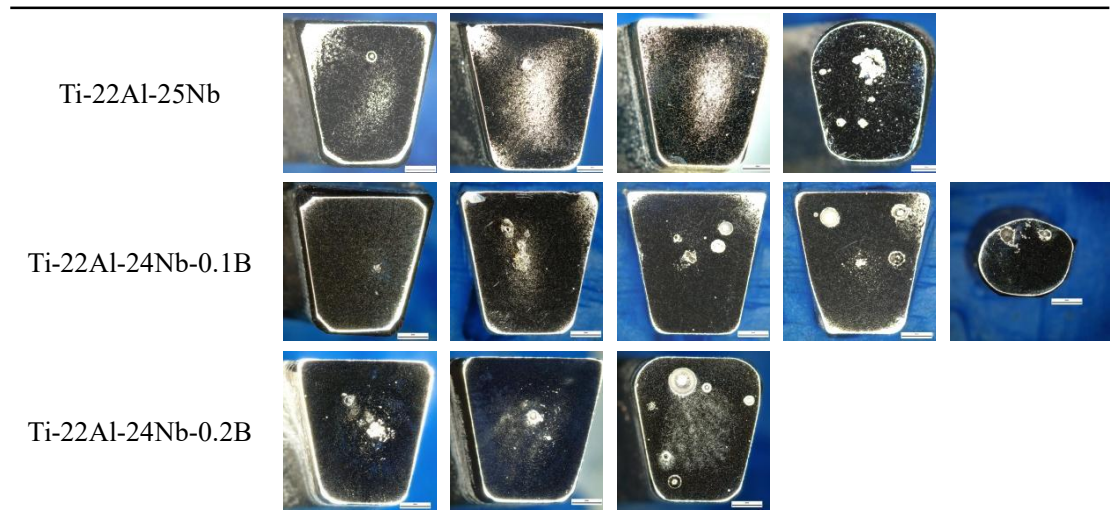

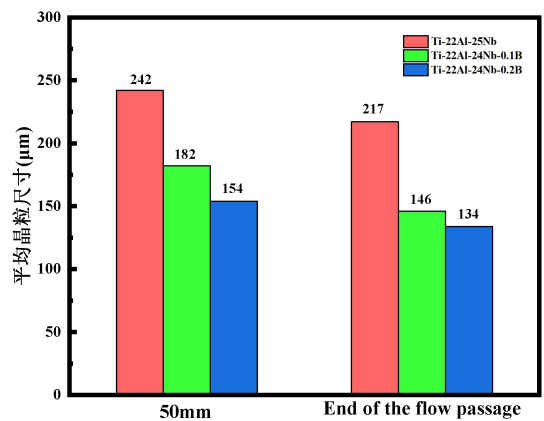

Fig.10 Average size of macro grains in cross sections of three Ti$_2$AlNb alloys at different positions from the sprue

The microstructure of the position closest to the main runner of the three alloys was shown in Fig.11. Dendritic characteristics can be observed on the microstructure of Ti-22Al-25Nb alloy, while Ti-22Al-24Nb-0.1B alloy and Ti-22Al-24Nb-0.2B alloy were not obvious. The reduction of Nb on the surface and the addition of trace B elements weaken the dendritic characteristics, and the finer the grain, the weaker the dendritic characteristics. The previous study [28] founded that the boride precipitated in the solid phase transformation process rather than in the solid-liquid two-phase region during the phase diagram calculation, which showed that the B element did not refine the grain through heterogeneous nucleation. According to the results of selected area electron diffraction, this kind of boride was determined as TiB (see Fig.12). The microstructure observation also shown that there are TiB-type boride precipitates on the grain boundary (see Fig.11(c)). These TiB phases could refine the grain by pinning the grain boundary to limit the grain growth rate in the B2 single-phase region. At the same time, the addition of B element reduced the surface tension in the liquid phase zone of Ti$_2$AlNb alloy melt, essentially reduced the critical nucleation work, and thus increased the probability and rate of grain nucleation. However, in the actual casting process, there was inevitable microsegregation, and TiB phase may also be preferentially crystallized in the melt as the core of heterogeneous nucleation of B2 grains, and the nucleation rate would be increased to refine the initial B2 grains. The refined grains would prevent the growth of dendrites into developed dendrite networks, inhibited the adverse effects of the increase in the solidification interval width on the fluidity, and reduced the flow resistance of liquid metal, which further improved the fluidity of the alloy [29, 30]. Based on this, the schematic diagram of Ti-22Al-25Nb alloy stopping flow mechanism was drawn as shown in Fig.13.

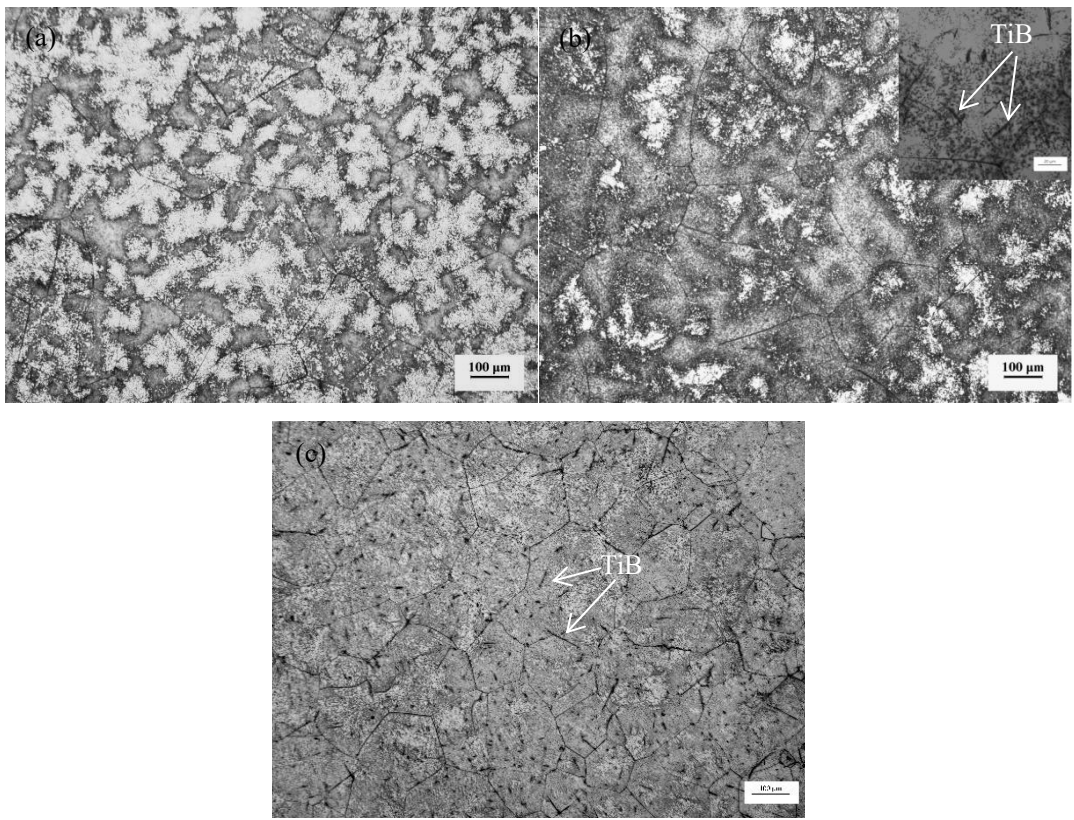

Fig.11 Optical microstructure of cross section of three Ti$_2$AlNb alloys at 5mm from the sprue
(a) Ti-22-25Nb alloy; (b) Ti-22Al-24Nb-0.1B alloy; (c) Ti-22Al-24Nb-0.2B alloy

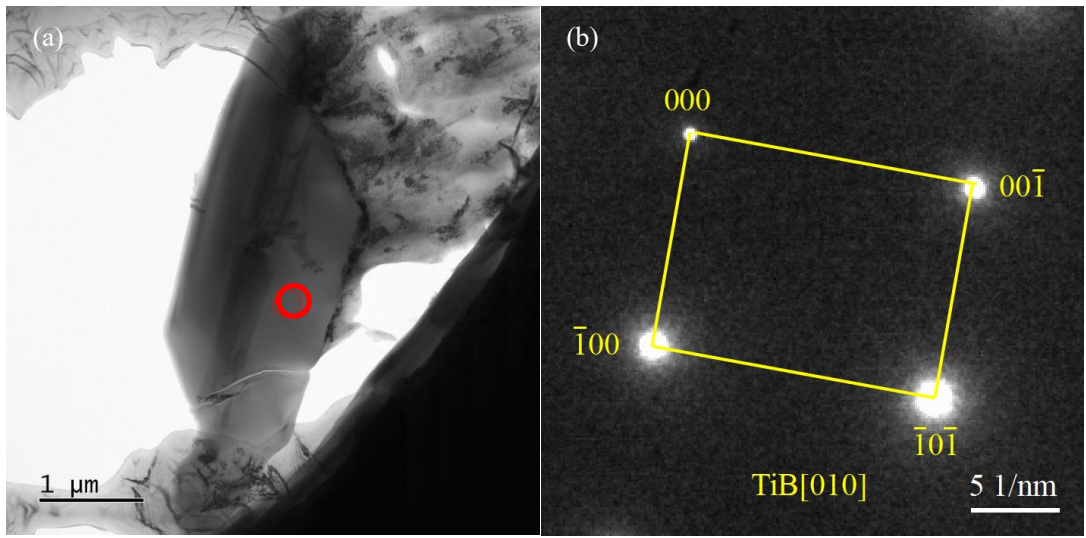

Fig.12 SAED results of Ti-22Al-24Nb-0.2B alloy
(a) Boride; (b) Diffraction spot

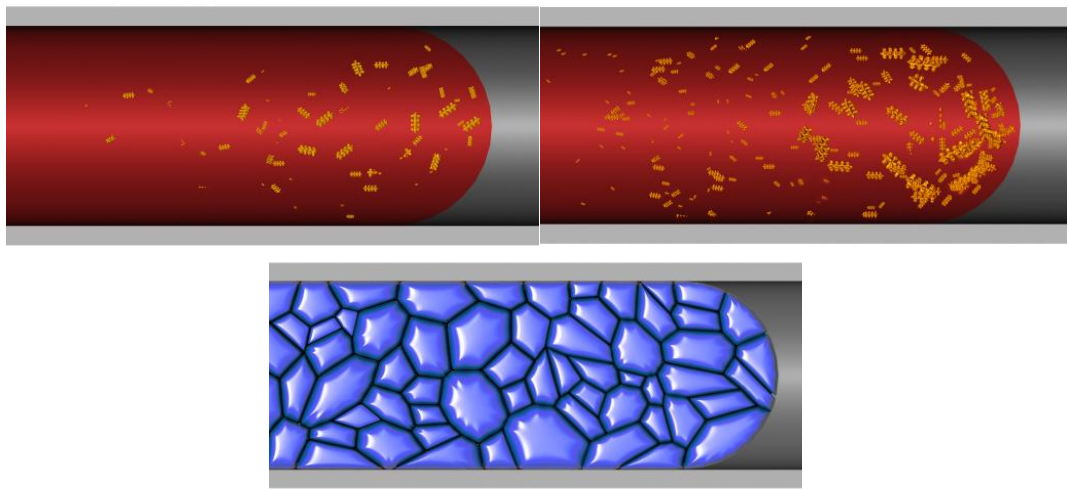

Fig.13 Schematic diagram of Ti$_2$AlNb alloy cessation mechanism

*4.2 Discussion on the mechanism of improving the fluidity of Ti$_2$AlNb alloy*

As shown in Fig.14, under the same pouring conditions, compared with Ti-22Al-25Nb alloy, the Superheat increment of Ti-22Al-24Nb-0.1B alloy was 8.9 °C, the solidification zone width increment was 16.8 °C. The melt thermal conductivity decreased, the viscosity decreased significantly, and the crystallization latent heat was equivalent in the whole liquid phase, but the thermal conductivity, dynamic viscosity, and crystallization latent heat of the solid-liquid two-phase zone increases by different degrees. Therefore, increasing the degree of superheat and the latent heat of crystallization were beneficial to improve the fluidity of the alloy, while increasing the width of solidification zone and dynamic viscosity were detrimental to improve the fluidity. It could be seen that the improvement effect of melt superheat, crystallization latent heat, viscosity and thermal conductivity reduction on the fluidity of Ti-22Al-24Nb-0.1B alloy was greater than the deterioration effect of its solidification zone width and two-phase dynamic viscosity increase on the fluidity. However, on the basis of Ti-22Al-24Nb-0.1B alloy, the thermal conductivity and crystallization latent heat above the solid point of Ti-22Al-24Nb-0.2B alloy had little change, and the dynamic viscosity of the whole liquid phase melt had little change. The superheat degree had only increased by 1.5 °C, but the solidification interval width had increased by 20.7 °C, and the dynamic viscosity of the solid-liquid two-phase zone had further increased. It could be seen that the increase of solidification zone width and dynamic viscosity of solid-liquid two-phase zone of Ti-22Al-24Nb-0.2B alloy had a significant impact on its fluidity deterioration.

The grain refinement of Ti-22Al-24Nb-0.2B alloy was greater, the developed dendrite network basically disappeared, and only a small amount of small size dendrites were dispersed in the B2 phase matrix. Increasing the superheat, increasing the latent heat of crystallization and refining the grain size should further improved the fluidity of the alloy, but the results of the fluidity verification test shown that the fluidity of Ti-22Al-24Nb-0.2B alloy was the worst of the three alloys, which indicated that the fluidity of the alloy may be affected by other factors. In aluminum matrix composites, the shape and volume fraction of reinforcement would affect the fluidity of metal matrix composite melt [31]. At a given temperature and volume fraction, the more irregular the shape of the reinforcement particles, the worse the fluidity of the composite melt. At a given temperature, the fluidity of the composite melt decreased with the increase of the volume fraction of the reinforcement particles. The borides in Ti-22Al-24Nb-0.1B alloy and Ti-22Al-24Nb-0.2B alloy were in long strip shape, but the volume fraction and size of borides in Ti-22Al-24Nb-0.2B alloy were significantly larger than that in Ti-22Al-24Nb-0.1B alloy [28]. From the previous analysis, it could be seen that this boride may precipitate before the liquid phase solidification, and act as a reinforcement similar to that in aluminum matrix composites. When the melt contained a large amount of irregular long borides, the melt viscosity in the solid-liquid two-phase region increased sharply, which may be the reason for the significant deterioration of the fluidity of Ti-22Al-24Nb-0.2B alloy.

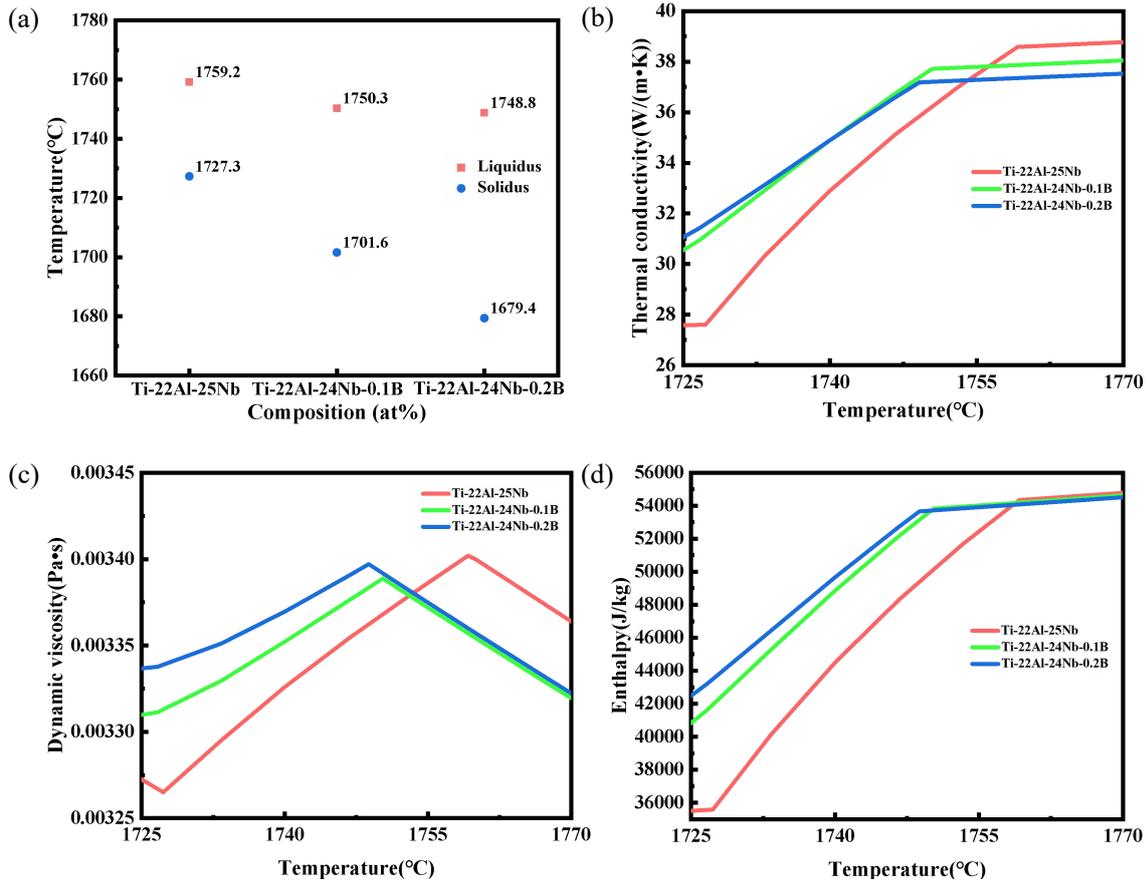

Fig.14 Thermophysical parameters of three Ti$_2$AlNb alloys vary with composition and temperature

5 Conclusion

In this paper, the effects of Al, Nb main elements, Fe, Mo, W, Co, B and Si trace elements and their contents on the fluidity of Ti-22Al-25Nb alloy were studied. The composition beneficial to improve the fluidity was selected through thermodynamic calculation of the thermophysical parameters that affected the fluidity, numerical simulation test and verification test of fluidity of the alloy. Finally, the improvement mechanism of the fluidity was discussed, the conclusions were as follows:

(1) According to the thermodynamic calculation results, it was speculated that for the improvement of fluidity, the appropriate reduction of Nb element was better than Al element; the addition of Fe、B and Si elements was conducive, the improvement effect of B element was best; the addition of Mo, W, Co was detrimental.

(2) The results of numerical simulation shown that reducing 1at.% Nb and adding 0.1~0.2at% B element was best for enhance of fluidity, which was better than addition of Fe and Si elements. Based on results of verification test, the composition for improving the fluidity best was Ti-22Al-24Nb-0.1B.

(3) Unlike conventional titanium alloys, the macrostructure of Ti-22Al-25Nb alloy and Ti-22Al-24Nb-0.1B alloy at different channel positions were equiaxed crystal structure, and the microstructure was characterized by fine dendrite. The research results shown that the stop flow mechanism of Ti$_2$AlNb alloy melt was the feature of width crystallization temperature range alloy.

(4) On the basis of Ti-22Al-25Nb alloy, the main reason for the fluidity improvement of Ti-22Al-24Nb-0.1B alloy was that the reduction of 1at% Nb and the addition of 0.1at% B increased the superheat and crystallization latent heat of the alloy, reduced the melt viscosity and thermal conductivity simultaneously. The fine grains prevented the growth of dendrites into a developed dendrite network, inhibited the adverse effect of the increase in the width of the solidification zone on the fluidity, it reduced the flow resistance of liquid metal. The reason why the addition of 0.2at% B did not actually improve the fluidity may be that the melt contains a large number of irregular and elongated TiB phases, which led to a sharp increase in the actual viscosity of the melt, thus hindering the flow of molten metal.